\documentclass[twocolumn]{aastex61}
\pdfoutput=1 
\usepackage{amsmath,amstext}
\usepackage[T1]{fontenc}
\usepackage{apjfonts} 
\usepackage[figure,figure*]{hypcap}
\usepackage{graphicx}
\usepackage{nicefrac}
\usepackage{color}
\usepackage{multirow}
\usepackage{tablefootnote}
\usepackage{epsfig}
\usepackage{xcolor}
\usepackage{hyperref}

\shorttitle{Estimate of the Accretion Disk Size in HE 0435-1223}
\shortauthors{Fian et al.}
\begin{document}
\title{Estimate of the Accretion Disk Size in the Gravitationally Lensed Quasar HE 0435-1223 using Microlensing Magnification Statistics}

\author{C. Fian}
\affiliation{Instituto de Astrof\'{\i}sica de Canarias, V\'{\i}a L\'actea S/N, La Laguna 38200, Tenerife, Spain}
\affiliation{Departamento de Astrof\'{\i}sica, Universidad de la Laguna, La Laguna 38200, Tenerife, Spain}
\author{E. Mediavilla} 
\affiliation{Instituto de Astrof\'{\i}sica de Canarias, V\'{\i}a L\'actea S/N, La Laguna 38200, Tenerife, Spain}
\affiliation{Departamento de Astrof\'{\i}sica, Universidad de la Laguna, La Laguna 38200, Tenerife, Spain}
\author{J. Jim{\'e}nez-Vicente}
\affiliation{Departamento de F\'{\i}sica Te\'orica y del Cosmos, Universidad de Granada, Campus de Fuentenueva, 18071 Granada, Spain}
\affiliation{Instituto Carlos I de F\'{\i}sica Te\'orica y Computacional, Universidad de Granada, 18071 Granada, Spain}
\author{J. A. Mu\~{n}oz}
\affiliation{Departamento de Astronom\'{i}a y Astrof\'{i}sica, Universidad de Valencia, E-46100 Burjassot, Valencia, Spain}
\affiliation{Observatorio Astron\'{o}mico, Universidad de Valencia, E-46980 Paterna, Valencia, Spain} 
\author{A. Hanslmeier}
\affiliation{Institute of Physics (IGAM), University of Graz, Universit{\"a}tsplatz 5, 8010, Graz, Austria}


\begin{abstract}
We present a measurement of the accretion disk size of the quadruple lensed quasar HE 0435-1223 from well-sampled 
13-yr COSMOGRAIL optical light curves. Using accurate time delays for the images A, B, C, and D, we modeled 
and removed the intrinsic quasar variability, and found microlensing events of amplitude up to 0.6, 0.4, and 0.5 mag 
in the images A, C and D respectively. From the statistics of microlensing magnifications in these images we use Bayesian methods to estimate the size of the quasar accretion disk. We have inferred the half-light radius 
for the accretion disk using two different methods, $R_{1/2} = 7.6_{-1.1}^{+12.0} \sqrt{M/0.3M_{\odot}}$ light-days 
(histogram product) and   $R_{1/2} = 7.7_{-1.2}^{+7.0} \sqrt{M/0.3M_{\odot}}$ light-days ($\chi^2$ criterion). The 
results are self-consistent and in good agreement with the continuum size predicted by single-epoch spectroscopy
and previous studies making use of narrow-band photometry of HE 0435-1223.
\end{abstract}
\keywords{gravitational lensing: micro -- quasars: individual (HE 0435-1223) -- accretion, accretion disks}

\section{Introduction \label{intro}}
The light curves of lensed quasar images provide a time history of the changes in brightness, and their analysis 
has important applications in cosmology (such as the determination of time delays to infer the Hubble constant, 
\citealt{Ref1964}, the estimate of peculiar velocities, \citealt{Mediavilla2016}, and in the study of quasar 
structure, \citealt{Chang1979,Chang1984}; see also \citealt{Kochanek2004} and \citealt{wam2006}). In this paper 
we  focus on the last application, using gravitational microlensing statistics to determine the quasar accretion 
disk size 
\citep{Pool2007,Morgan2010,Sluse2011,Blackburne2011,Motta2012,Blackburne2014,Blackburne2015,Mosquera2011a,Jimenez2012,Jimenez2014,Hainline2013,Mosquera2009,Mosquera2013,Jimenez2015b,Jimenez2015a,Mediavilla2015,Munoz2015,fian2016}. 
The term ``microlensing'' describes the flux variations produced by stars in the foreground lens galaxy that 
result in magnification (or demagnification) of the multiple quasar images (\citealt{Chang1979,Kochanek2004,Blackburne2014}; 
see also the review by \citealt{wam2006}). These flux variations between images are not correlated as  would be expected from 
intrinsic quasar variability. The magnification produced by microlensing depends strongly upon the angular size of the source, 
with smaller emission regions showing larger flux anomalies and larger sources smoothing out the light curves (\citealt{Mosquera2011b,Blackburne2011,Blackburne2014}).\\

We study the light curves of the quadruple lensed quasar HE 0435-1233 discovered during the
Hamburg/ESO Survey (HES) in the southern hemisphere (\citealt{Wisotzki2002}). The lens lies in a group of 
galaxies of at least 12 members (\citealt{Sluse2017}). At a redshift of $z_s$ = 1.689 (\citealt{Wisotzki2002}), HE 0435-1223 is 
lensed by a foreground galaxy at a redshift of $z_l$ = 0.455 (\citealt{Eigenbrod2006}) into four bright point
sources (plus a fuzzy object in the center) in a nearly symmetric cross-shaped configuration (\citealt{Wisotzki2002,Blackburne2014}). 
The maximum separation between images is 2.6$''$ (\citealt{Wisotzki2002}) and their time delays are relatively small owing to
the symmetric distribution of the images around the lensing galaxy (\citealt{Mosquera2011a}).\\

After correcting for the time delays (components A and C lead, followed by the saddlepoints B and D,
see \citealt{Wisotzki2002}) and mean magnitude differences between the images, we find clear indications
of microlensing flux variability in the residuals of the light curves (i.e., the differences between the 
observed light curves and the modeled intrinsic variability of the quasar).  \\

Our aim is to estimate from the statistics of microlensing the size of the accretion disk of the lensed quasar. 
We use flux ratios of a large enough source in the quasar so as to be insensitive to microlensing in order to 
establish the baseline for no microlensing magnification (e.g., \citealt{Mediavilla2009}), from which the 
amplitude of the microlensing magnification can be measured. We then compare the histogram of microlensing 
magnifications obtained from the observations (corresponding to the monitoring time interval) with the simulated 
predictions of microlensing variability for sources of different sizes (\citealt{fian2016}). This comparison 
 allows us to evaluate the likelihood of the different values adopted for the size. In this way we extend the 
single epoch method to all the epochs in the available 13-yr light curves, thereby increasing the statistical significance. 
We  use the optical light curves obtained from the COSMOGRAIL\footnote{The data has been made publicly available by the COSMOGRAIL collaboration through \url{https://cosmograil.epfl.ch/}.} project (\citealt{Bonvin2017}) to infer 
microlensing flux variability and the radio data from \citealt{Jackson2015} to estimate the baseline for no microlensing 
variability. In the present study we improve the methods for obtaining the accretion disk size discussed in 
\citealt{fian2016} by using more realistic estimates of the scatter of the modeled histograms.\\

The paper is organized as follows. In Section 2 we present the COSMOGRAIL light curves of each image of HE 0435-1223. 
In Section 3 we model the intrinsic variability and examine the flux ratios between the images. We outline our approaches to compute the accretion disk size in Section 4. 
Section 5 is devoted to analyzing our results and comparing them with past estimates. In Section 6 we will discuss the impact of errors and uncertainties on the size estimates. Finally, in Section 7 we briefly conclude our results and discuss future perspectives.

\section{Data}
\begin{figure*}
\centering
\includegraphics[width=18cm]{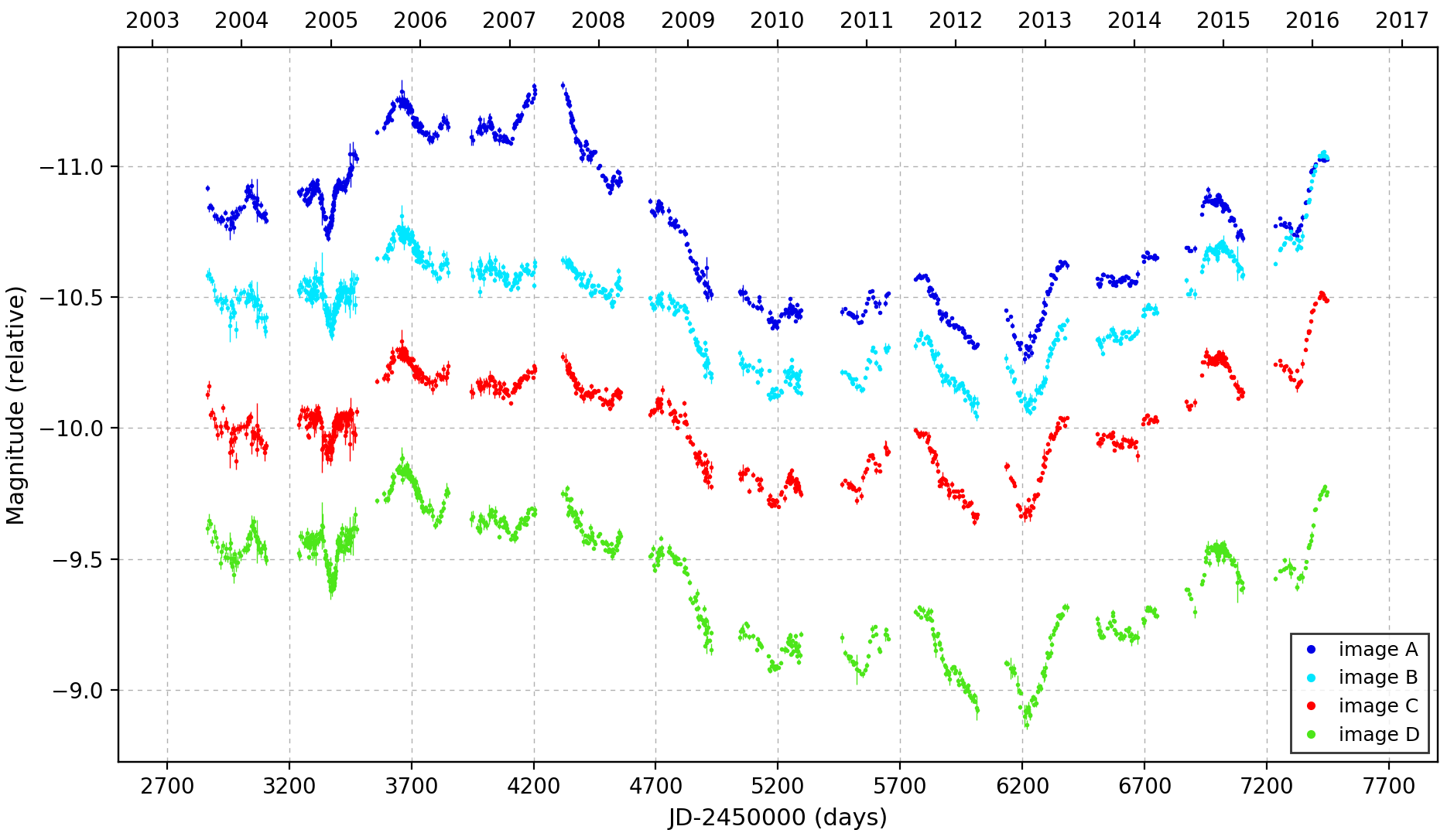}
\caption{Light curves of the four lensed images A, B, C, and D of the quasar HE 0435-1223 from  2003 August
to  2016 February as obtained by the COSMOGRAIL project (see \citealt{Bonvin2017}). Horizontal axes show 
the Julian (bottom) and Gregorian (upper) dates. The light curves of the images B, C, and D are shifted by 
$-0.2$ mag, 0.3 mag, and 0.6 mag, respectively, so that they do not overlap with each other.\\}
\label{lightcurve}
\end{figure*}

The fluxes of the four images of HE 0435-1223 were monitored from  2003 August until  2016 February
in the optical $R$-band as a part of the COSMOGRAIL program. The data set consists of 884 epochs (i.e., 884 
nights) and the average sampling rate is once every fifth day. Figure \ref{lightcurve} shows the 13-yr 
light curves of  images A--D of HE 0435-1223 as published in figure 2 of \citealt{Bonvin2017}. The relative 
shifts in magnitude between the images are chosen to ease visualization. The similarity among the four 
well-sampled light curves is immediately noticeable, although it can be seen that they would not overlap 
perfectly when shifted in time and magnitude. This mismatch between the light curves is interpreted as 
microlensing caused by stars in the lensing galaxy. Quasars are time variable, making it necessary to 
separate microlensing from intrinsic variability by modeling and removing the later one.\vspace{1cm}

\section{Intrinsic Variability and Microlensing}
The images of multiple lensed quasars arrive with relative delays of hours up to years because of 
the different paths taken by their light. Intrinsic variability of the source coupled with the light 
path time delay between the quasar images can mimic flux ratio anomalies. To correct for this, we use the 
time delay estimates of \citealt{Bonvin2017} and shift the light curves by $\Delta t_{AB} = -8.8$ days, 
$\Delta t_{AC} = -1.1$ days, and $\Delta t_{AD} = -13.8$ days. Owing to the symmetry of the image configuration 
in this system, the time delays are very short, meaning that intrinsic variations (assumed to be much slower)
will show up quasi-simultaneously in all four images.\\

\begin{figure*}[h]
\centering
\includegraphics[width=17.5cm]{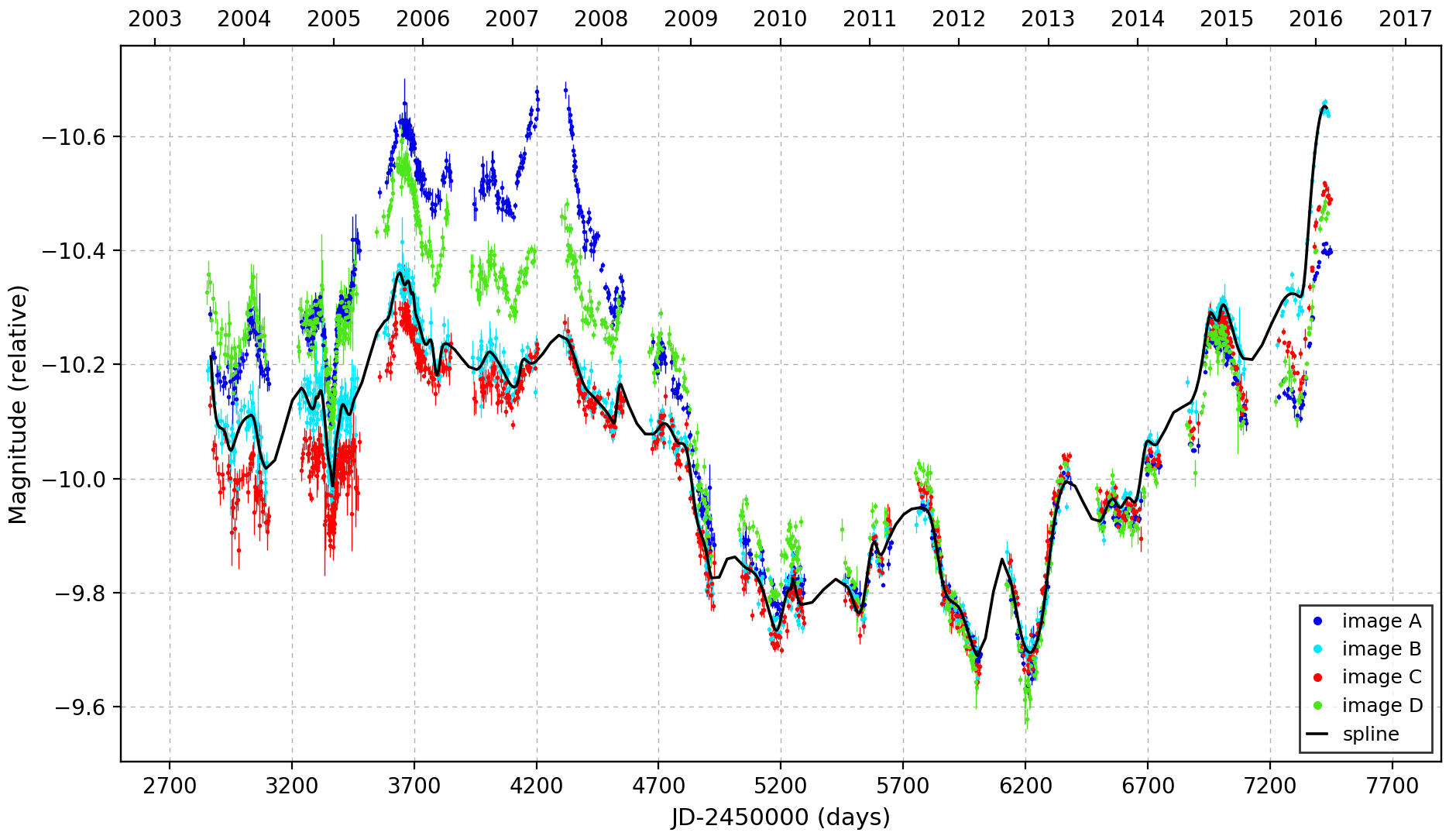}
\caption{Image A, B, C, and D light curves of HE 0435-1223 in their overlapping region after shifting by 
the respective time delays (and magnitude differences). The model of the intrinsic variability of
the quasar (spline-fitted to the light curve B) is shown in black.\\}
\label{spline}
\centering
\includegraphics[width=17.3cm]{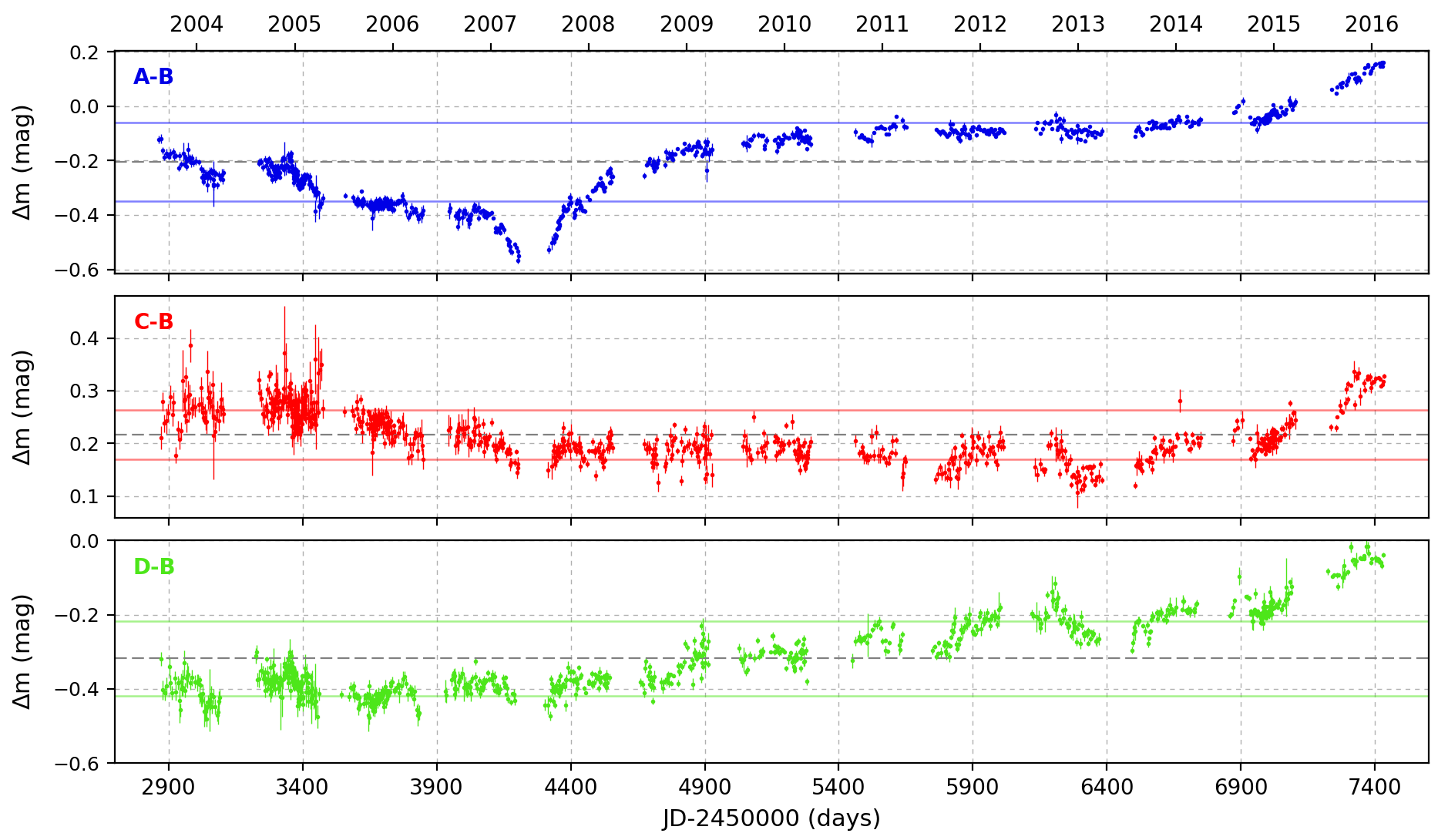}
\caption{Differential microlensing variability of the light curves A, C, and D compared to a spline fit
to the light curve B. The dashed horizontal lines show the mean value of the residuals. The residual 
magnitudes clearly show that microlensing is present in  light curve A.}
\label{residuals}
\end{figure*}

After shifting the light curves in time and correcting for the magnitude difference between the images, 
we perform a single spline fit to the B light curve in order to model the intrinsic variability of the quasar.
We assume that the flux variations in image B are mainly intrinsic, as several authors claim that the B 
image is the least affected by stellar microlensing (\citealt{Courbin2011,Motta2012}), whereas A and D are 
affected by strong microlensing variations (\citealt{Wisotzki2003,Kochanek2006,Courbin2011,Mosquera2011a,Ricci2011,Motta2012}). 
Figure \ref{spline} shows the simulated quasar variability (black solid line) with the A--D light curves 
superimposed. We obtain a source variability of $\sim$1 mag. Although we use the spline fit to make a 
reasonable estimate for the intrinsic variability of the source, some contribution from microlensing 
variability in image B is likely present.\footnote{In any case, this is irrelevant because in our treatment we also consider the
contributions of the B image to microlensing in the simulated difference light curves.} We subtract the spline from the
light curves and obtain the microlensing difference light curve as follows: 
$\Delta m_X = m_X - m_{Bfit} - (m_X-m_B)_{radio}$ where X = A, C, D, assuming that the ratios 
between the radio data from \citealt{Jackson2015} (36.0$\pm$2.1 $\mu$Jy for A, 26.4$\pm$2.1 $\mu$Jy 
for B, 34.3$\pm$2.1 $\mu$Jy for C, and 16.1$\pm$2.1 $\mu$Jy for D) represent the true magnification ratios 
of the images in the absence of microlensing. The radio-emitting regions of quasars should provide a good
estimate of the real magnification ratios of the images as they are supposed to arise from a large enough 
region so as not to be affected by microlensing (see \citealt{Mediavilla2009}). In the three panels in
Figure \ref{residuals} the microlensing residuals of the A, C, and D light curves after subtraction of 
the spline fit are shown. Microlensing variability can be seen, particularly in the A--B residual light 
curve, where image A seems to have been undergoing a microlensing event between the fourth and fifth seasons,
whereas C and D remained mainly constant.\\

\section{Bayesian Source Size Estimation}
The effect of finite source size is to smooth out the flux variations in the light curves of lensed quasars 
caused by stars in the galaxy. Hence, microlensing is sensitive to the size of the source (\citealt{Morgan2010}, 
see also the review by \citealt{wam2006}), and we  use quantitative Bayesian methods together with our 
determinations of microlensing magnification amplitude to estimate the accretion disk size in the HE 0435-1223 lensed quasar.\\

\subsection{Simulated Microlensing Histograms}
We simulate the microlensing of a finite-size source using microlensing magnification maps created with the Inverse 
Polygon Mapping method described by \citet{Mediavilla2006,Mediavilla2011}. Each map (appearing as a network of 
high-magnification caustics separated by regions of lower magnifications) corresponds to a specific quasar 
image and shows the microlensing magnification at a given source position. The general characteristics of 
the magnification maps are determined (for each quasar image) by the local convergence, $\kappa$, and the local 
shear, $\gamma$, which were obtained by fitting a singular isothermal sphere with an external shear (SIS+$\gamma_e$), 
such as might be generated by the tide from a neighboring galaxy or cluster, to the coordinates of the images. The 
local convergence is proportional to the surface mass density and can be divided into 
$\kappa = \kappa_c+\kappa_\star$, where $\kappa_c$ is the convergence due to continuously distributed matter 
(e.g., dark matter) and $\kappa_\star$ is due to the stellar-mass point lenses (e.g., microlens stars in the galaxy). 
The values of $\kappa$ and $\gamma$ (taken from \citealt{Mediavilla2009}) are listed in Table \ref{lensmodel}.
We use a surface mass fraction in stars $\kappa_{*}$ of 10$\%$ \citep{Mediavilla2009} and generated 2000$\times$2000 pixel 
magnification maps with a size of 19.3$\times$19.3 Einstein radii$^2$. We get a resolution of 0.2 light-days per pixel, which is much smaller than the size of the optical accretion disk of the quasar. The value of the Einstein radius
for this system is $2.84 \times 10^{16} \sqrt{M/0.3 M_\odot}$ cm = $11 \sqrt{M/0.3 M_\odot}$ light-days at the lens
plane (\citealt{Mosquera2011b}). We randomly distribute stars of mass $M = 0.3\ M_\odot$ across the microlensing
patterns to create the microlens convergence $\kappa_\star$. The source sizes can be scaled to a different 
stellar mass, $M$, using $r_s \propto \sqrt{M}$. The ratio of the magnification in a pixel to the average 
magnification of the map gives the microlensing magnification at the pixel and histograms of maps normalized 
to the mean deliver the relative frequency of microlensing magnification amplitude for a pixel-size source.\\

\begin{figure*}
\centering
\includegraphics[width=18cm]{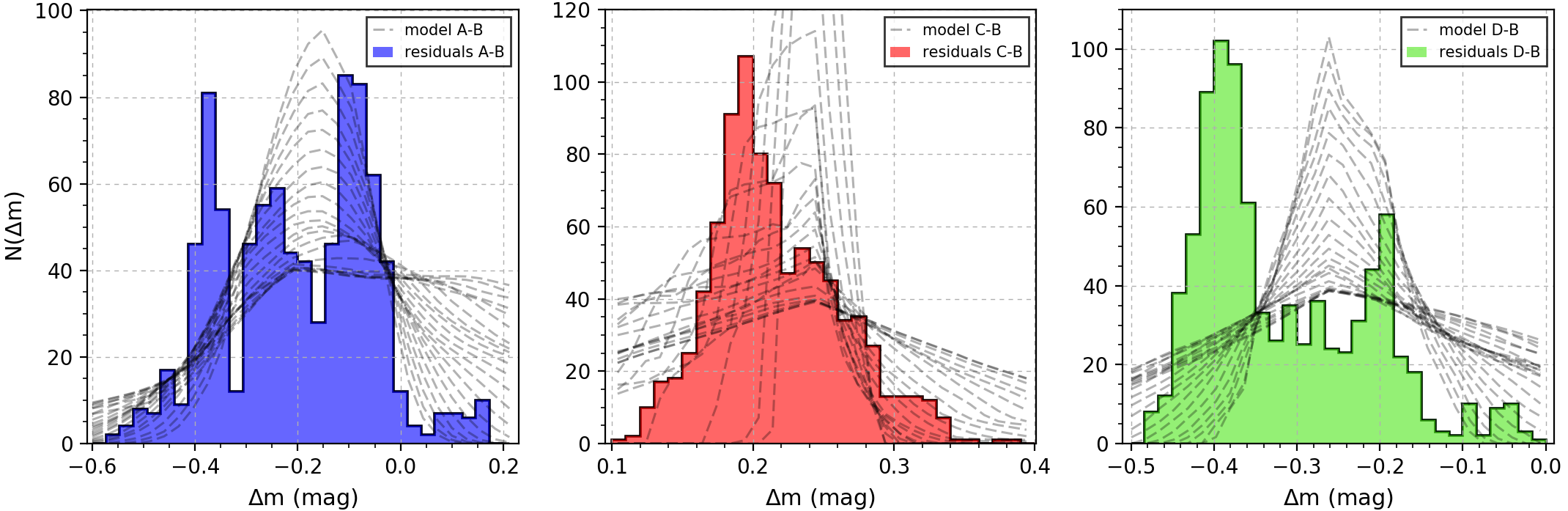}
\caption{Microlensing frequency distributions obtained from the observed light curves (histograms) and the 
simulated microlensing magnification maps (dashed lines). The polygonal lines present model histograms for 
different values of $r_s$ (from 0.5 to 21.5 light-days).\\}
\label{histogram}
\end{figure*}

\begin{table}
\centering
\tabcolsep=0.3cm
\renewcommand{\arraystretch}{0.85}
\caption{Lens Model Parameters}
\begin{tabular}{ccc} \hline \hline \vspace*{-3.5mm}\\
Image & $\kappa$ & $\gamma$ \\ \hline 
A & 0.46$\pm$0.03 & 0.39$\pm$0.04 \\ 
B & 0.52$\pm$0.12 & 0.59$\pm$0.06 \\
C & 0.46$\pm$0.05 & 0.39$\pm$0.08 \\
D & 0.56$\pm$0.14 & 0.64$\pm$0.08 \\ \hline 
\end{tabular}
\label{lensmodel}    
\end{table}
To model the structure of the unresolved quasar source we considered a circular Gaussian intensity profile of 
size $r_s$, $I(R) \propto exp(-R^2/2 r_s ^2)$. It is generally accepted that the specific shape of the source's 
radial emission profile is unimportant for microlensing flux variability studies because the results are 
essentially controlled by the half-light radius rather than by the detailed source profile \citep{Mortonson2005}. 
The characteristic size $r_s$ is related to the half-light radius; that is, the radius at which half of the light 
at a given wavelength is emitted, by $R_{1/2} = 1.18\ r_s$. Finally, we convolve the magnification maps with 
Gaussians of 22 different sizes over a linear grid which spans from $r_s =$ 0.5 to 22.5 light-days. The movement
of a large source across the magnification map is equivalent to a point source moving across a version of the map 
that has been smoothed by convolution with the intensity profile of the source. Strong anomalies are evidence for 
a relatively small source, whereas low microlensing magnifications could be due to a large source size or to a 
location of the source in a relatively calm region of the magnification map. After convolution we normalized each 
magnification map by its mean value. The histograms of the normalized map represent the histograms of the 
expected microlensing variability. Thus, we obtain 22 different microlensing histograms corresponding to different 
source sizes for each of the images A, C, and D. Finally, cross-correlating the histograms of A, C, and D with the 
histogram of B, we built the microlensing difference histograms A-B, C-B and D-B for different values of $r_s$ to be 
compared with the experimental histograms obtained from the observed light curves (see Figure \ref{histogram}).\\

\subsection{Observed Microlensing Histograms}
From the residuals that represent the differential (with respect to B, the image least prone to microlensing) 
microlensing of the A, C, and D images (see Figure \ref{residuals}), we have obtained the microlensing histograms; 
i.e., the frequencies in which each microlensing amplitude appears in the microlensing light curves. We adopted 
a bin size of 0.05 mag. In Figure \ref{histogram} we compare the A-B, C-B, and D-B modeled magnification
histograms corresponding to convolutions with sources of different values of $r_s$ (dashed lines) with the 
experimental microlensing histograms. Large values of $r_s$ smear out the network of microlensing magnification
caustics and reduce their dynamic range, thereby causing the histograms to become narrower.\\  

\subsection{Methods \label{methods}}
\begin{figure}
\centering
\includegraphics[width=8.7cm]{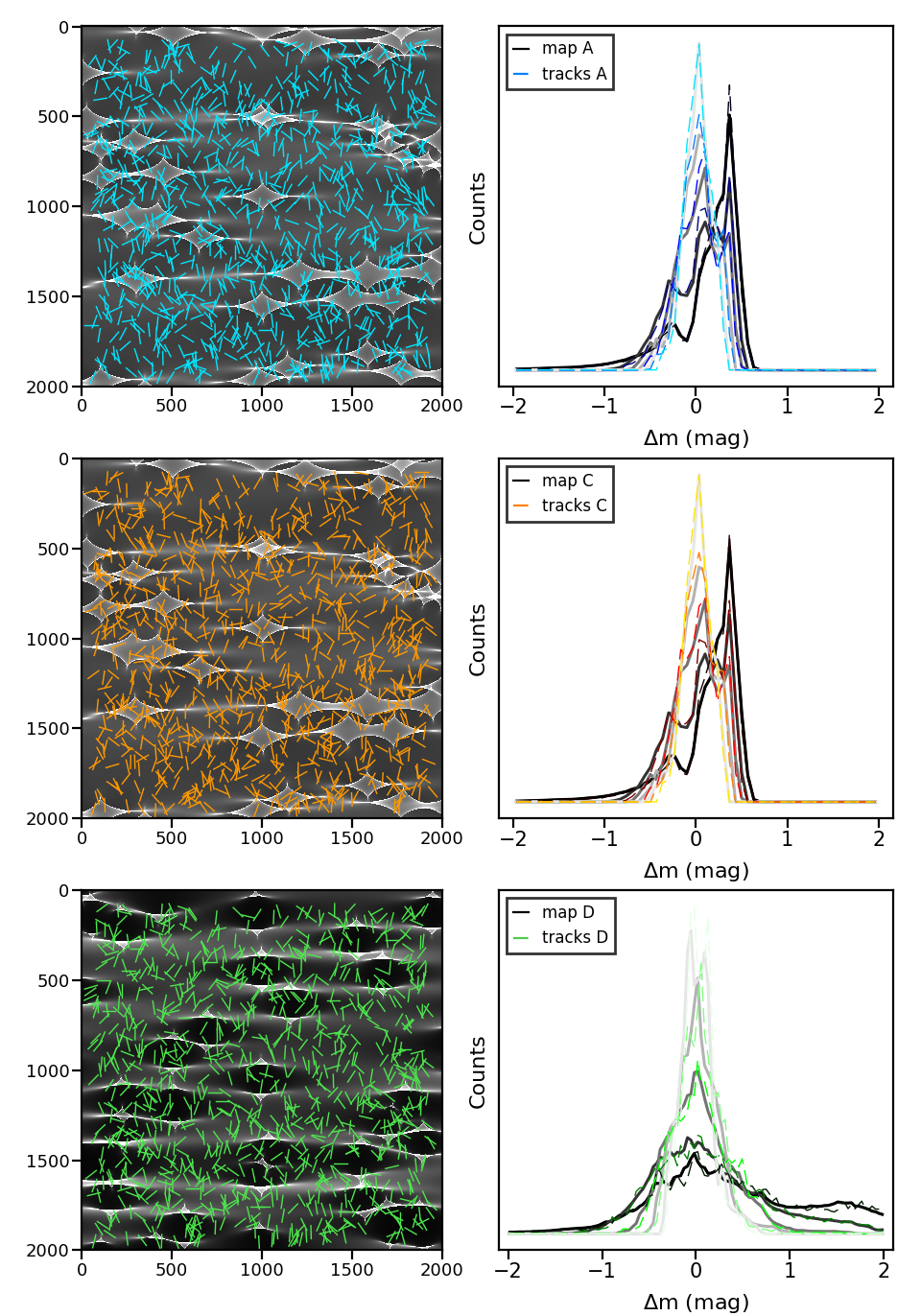}
\caption{Random tracks superposed on magnification maps of images A (top left), C (middle left), and D (bottom left). 
The gray scale shows the unconvolved map with lighter colors indicating higher magnifications. The lines show the 
source trajectories across the pattern for the COSMOGRAIL monitoring period. Histograms derived from the magnification
maps (gray) and the tracks for the images A (blue), C (red), and D (green) are shown in the right-hand panels. Different 
color shadings stand for different convolutions with $r_s$. Positive numbers of magnification denote demagnification.}
\label{tracks}
\end{figure}

To study the likelihood of the different $r_s$ values we compare the microlensing histograms inferred from the model 
for different values of $r_s$ with the histograms of the data using two different statistics: 

\begin{enumerate}
\item[(a)] Histogram product\footnote{We propose heuristically this statistics based in the distance between histograms, related to the Pearson's correlation coefficient.}, defined as 
\begin{equation}
P_X(r_s) = \sum_{i=1}^{N_{bin}} h_{X-B}^{i}\ \hat{h}_{X-B}^{i}(r_s),
\end{equation}
where $h_{X-B}^{i}$ and $\hat{h}_{X-B}^{i}(r_s)$ are the observed and modeled histograms, and $N_{bin}$ is 
the number of bins. This is a natural extension of the single-epoch method. After multiplying the probability 
distributions corresponding to A, C, and D we obtain the PDF of the source size,
\begin{equation}
P(r_s) = P_A(r_s) \cdot P_C(r_s) \cdot P_D(r_s).
\label{prs}
\end{equation}


\item[(b)] Pearson's $\chi^2$-test. This is a test suited to measure the ``distance'' between two histograms. 
After normalizing all simulated histograms to the number of counts in the real data, we measure the goodness 
of fit between the histograms inferred from the model (for different source sizes $r_s$) and the histogram of 
the observed microlensing differences with a $\chi^2$- statistic:
\begin{equation}
\chi^2 = \sum_{i=1}^{N_{bin}} \frac{(h_{X-B}^{i}-\hat{h}_{X-B}^{i}(r_s))^2}{(\sigma_{X-B})^{2}} 
\label{ns}
\end{equation}
with 
\begin{equation}
\sigma_{X-B} = \sqrt{{(\sigma_{X-B}^{model})^{2}+(\sigma_{X-B}^{obs})^{2}}},
\end{equation}

where $\sigma_{X-B}^{model}$ and $\sigma_{X-B}^{obs}$ are the uncertainties of the model and the observations, and 
\begin{equation}
P_X(r_s) \propto e^{-\frac{\chi^2}{2}}.
\end{equation}
Multiplying them together, we obtain the PDF of the source size,
\begin{equation}
P(r_s) = \prod\limits_{X} P_X(r_s).
\end{equation}

In our previous paper (\citealt{fian2016}) we noticed that the application of the $\chi^2$ method led to
underestimation of the uncertainties in the size. We obtained relatively high values for $\chi^2$, which indicates 
that we were probably underestimating the intrinsic scatter of the model histograms obtained from the magnification
maps. It should be taken into account that the observed light curves correspond to a tiny track on the magnification 
map and that the scatter between the histograms corresponding to different tracks can be large. In other words, we 
need to know the scatter of a simple track realization with respect to the mean. In this work we propose to control 
this problem by estimating the uncertainties as follows:
Once we have the convolved magnification maps for each image, we run 1000 tracks across them, at random starting points 
and in random directions (of time length corresponding to the observed light curve) in order to estimate the scatter in the 
magnification histogram. Scaling the COSMOGRAIL monitoring period of HE 0435-1223 (13 yr) with the Einstein crossing time computed by \citealt{Mosquera2011b}, we estimated the distance in pixels traveled by the source along the caustic pattern. For each image 
and each convolved map, we build histograms of the set of tracks and compute their average. From these average histograms
we estimate the dispersion in each bin to get the uncertainty contributions for each image that we will use in quadrature 
for our $\chi^2$ calculations. Coming from different regions of the magnifications map, the scatter among the histograms 
of the random trajectories is high. We calculated the $\chi^2$ for 22 source sizes $r_s$, spaced linearly between
0.5 and 21.5 light-days. The minimum $\chi^2$ using this estimate of the histogram uncertainties is $\sim$2 for 
A and C, and $\sim$6 for D, respectively.\\

In Figure \ref{tracks} we show the random source trajectories from which we build these histograms superposed on the 
magnification map for each image (left panels). In order to make the caustics and cusps more easily visible, in this 
figure we did not convolve the magnification maps with the source size. In the right-hand panels of Figure \ref{tracks} 
the histograms of the whole map are shown in gray and the average histograms of the tracks are shown in color, with
different shadings standing for different convolutions of $r_s$.

\end{enumerate}

\section{Results and Discussion}
\begin{figure}
\centering
\includegraphics[width=8.5cm]{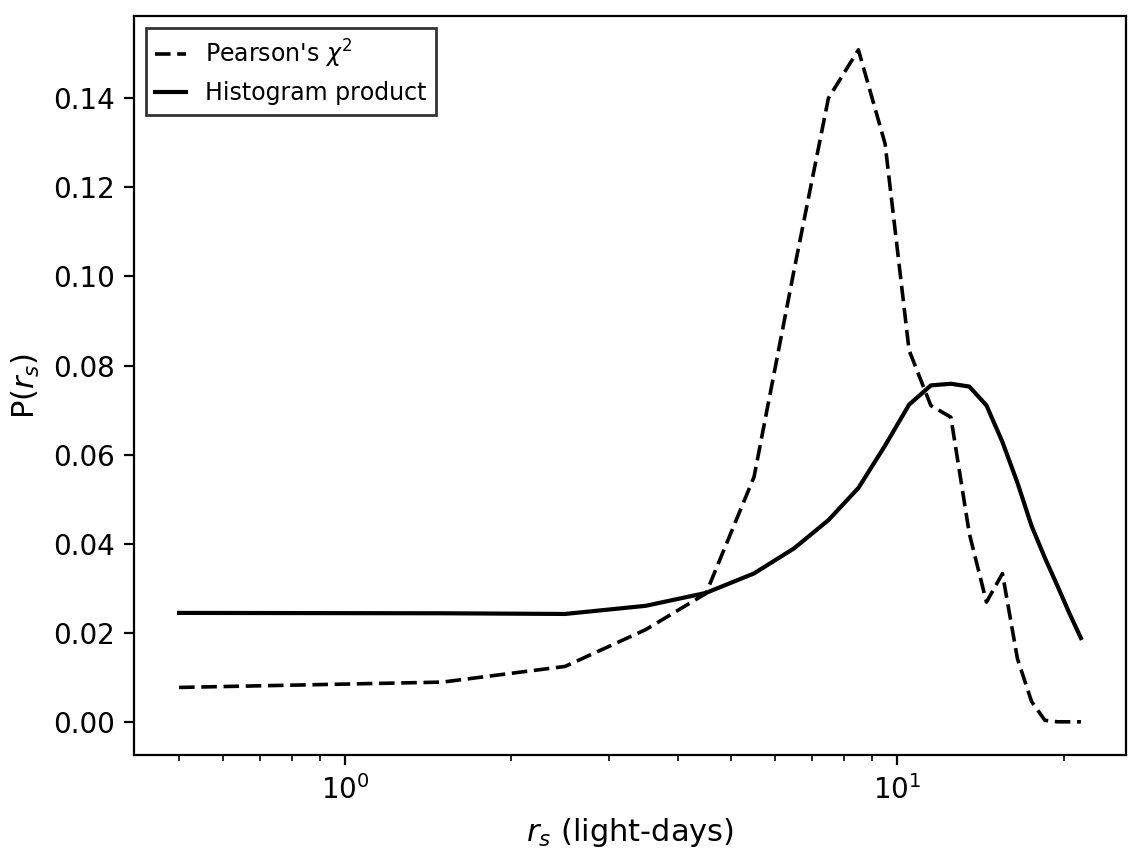}
\caption{Probability distributions of the source size $r_s$ for the  histogram-product (solid line) and  Pearson's $\chi^2$ (dashed line).}
\label{PDFs}
\end{figure}

The resulting normalized probability distributions obtained using the methods discussed in Section \ref{methods} can 
be seen in Figure \ref{PDFs}. Using a logarithmic prior, we found a size of the region emitting the $R$-band emission of 
$\langle r_s \rangle = 7.1^{+9.4}_{-1.6} \sqrt{M/0.3M_{\odot}}$ for 68$\%$ confidence estimates for the histogram product 
(solid line). Using  Pearson's $\chi^2$ statistic, we predict a source size of 
$ \langle r_s \rangle = 7.5^{+5.0}_{-1.0} \sqrt{M/0.3M_{\odot}}$ light-days (dashed line). We obtained values of 
$\chi^2$ $\sim$2 for A and C, and $\sim$6 for D.\\ 

Multiplying by a factor of 1.18, we convert $r_s$ to half-light radii. Our result for each method expressed
in terms of the half-light radius, $R_{1/2} = 8.4^{+11.1}_{-1.9} \sqrt{M/0.3 M_{\odot}}$ light-days (histogram 
product) and $R_{1/2}=8.9^{+5.9}_{-1.2} \sqrt{M/0.3 M_{\odot}}$  light-days (Pearson's $\chi^2$) is in good 
agreement with the estimates by \citealt{Motta2017} ($R_{1/2} = 19^{+8}_{-6} \sqrt{M/0.3 M_{\odot}}$ light-days), 
\citealt{Mosquera2011a} ($R_{1/2} = 5^{+4}_{-4} \sqrt{M/0.3 M_{\odot}}$ light-days) and \citealt{Blackburne2011} 
($R_{1/2} = 6.7^{+3.0}_{-2.5} \sqrt{M/0.3 M_{\odot}}$ light-days) for this system. All the estimates have been scaled to a $\lambda_0 = 2417$ \AA\ using $R_{1/2} = (\lambda_0/\lambda)^p\ R_{1/2}(\lambda)$. Our estimates for the size are 
also in good agreement with the average determinations obtained for a sample of lensed quasars by 
\citet{Jimenez2012,Jimenez2014,Jimenez2015a,Jimenez2015b} when a fraction of mass in stars of $10\%$ is considered.\\

\section{Impact of Uncertainties on Size Estimates}\color{black}

\begin{figure*}
\centering
\includegraphics[width=16cm]{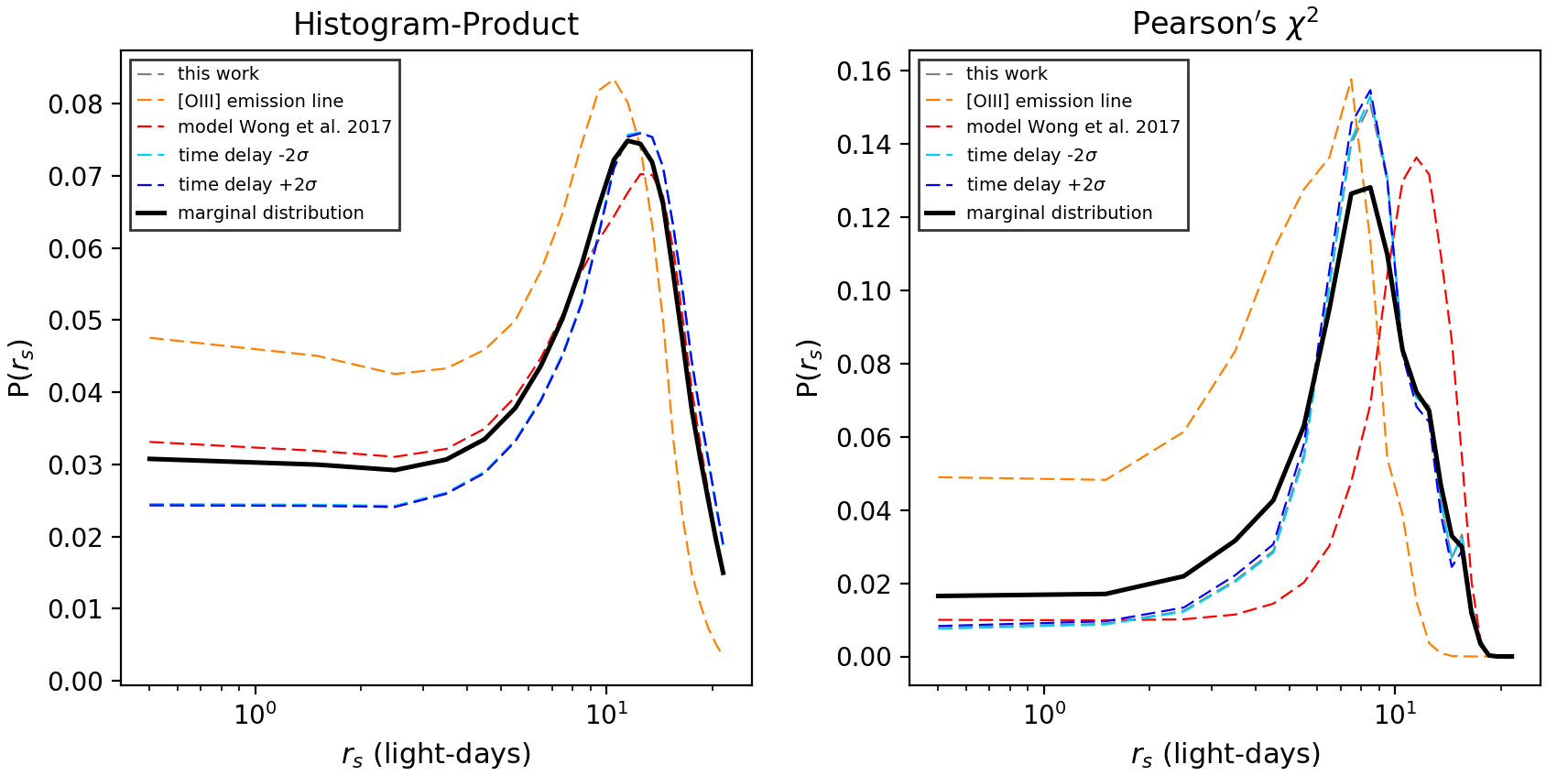}
\caption{Probability distributions of the source size $r_s$ for the histogram-product (left) and Pearson's $\chi^2$ (right) for various models/data-related analysis (dashed lines), as indicated by the legend. Notice that the probability distributions of Figure \ref{PDFs} overlap with those derived from the shifted $\pm$2$\sigma$ time delays. The thick black line represents the sum of all the distributions, which accounts for the different systematic and model dependent uncertainties.\\}
\label{marginal}
\end{figure*}

In Table \ref{errors} and Figure \ref{marginal} we summarize the impact of the different sources of uncertainty on the size estimates. The relatively small uncertainties in the time delays (less than one day, see Table \ref{time}) do not induce significant changes in the disk size. The effect of microlensing on the time delays (see \citealt{Tie2018}) is smaller than the uncertainties of the modeled time delays and has not influence on the size either. We studied the change of the size when we use the narrow line flux ratio measurements by \citealt{Nierenberg2017} as a baseline for no microlensing instead of the radio measurements by \citealt{Jackson2015} (see Table \ref{narrow}). We obtain $\sim30\%$ smaller sizes for the histogram-product ($\sim50\%$ for the Pearson's $\chi^2$) using the $[$OIII$]$ emission line (note that the narrow line cores could be affected by extinction). We checked the robustness of our results with respect to the macromodel by comparing it with the parameters (convergence and shear, see Table \ref{macro}) inferred from the inverse magnification tensor in \citealt{Wong2017}, where the authors explicitly model the effect of nearby perturbers. After recomputing the magnification maps we repeat all calculations, obtaining similar values for the half-light radius of the accretion disk ($R_{1/2} = 7.3_{-1.2}^{+12.2} \sqrt{M/0.3M_\odot}$ light-days for the histogram-product and $R_{1/2} = 10.3_{-1.3}^{+8.4} \sqrt{M/0.3M_\odot}$ light-days for Pearson's $\chi^2$). We marginalized over all distributions listed in Table \ref{errors} and obtained a half-light radius of $R_{1/2} = 7.6_{-1.1}^{+12.0}\sqrt{M/0.3M_\odot}$ light-days for the histogram-product. Consistent results are obtained with Pearson's $\chi^2$ ($R_{1/2} = 7.7_{-1.2}^{+7.0} \sqrt{M/0.3M_\odot}$ light-days). Notice that the large asymmetry in the uncertainties on $r_s$ arise from the progressive lack of sensitivity of microlensing to changes in the size when the size increases.

\begin{table}
\centering
\tabcolsep=0.25cm
\renewcommand{\arraystretch}{1}
\caption{Half-Light Radius $R_{1/2}$ in $\sqrt{M/0.3M_\odot}$ light-days.}
\begin{tabular}{lcc} \hline \hline \vspace*{-4.5mm}\\
Source & Histogram-product & Pearson's $\chi^2$ \\ \hline 
Time delay -2$\sigma$ & $8.4_{-1.9}^{+11.1}$ & $8.9_{-1.9}^{11.1}$\\ 
Time delay +2$\sigma$ & $8.4_{-1.2}^{+5.9}$ & $8.6_{-0.9}^{+6.1}$ \\
$[$OIII$]$ Emission Line & $5.9_{-1.8}^{+10.0}$ & $4.7_{-0.6}^{+5.3}$ \\
Model \citealt{Wong2017} & $7.3_{-2.0}^{+12.2}$ & $10.3_{-1.3}^{+8.4}$ \\ 
Marginal Distribution & $7.6_{-1.1}^{+12.0}$ & $7.7_{-1.2}^{+7.0}$ \\
\hline 
\end{tabular}
\label{errors}    
\end{table}

\begin{table}
\centering
\tabcolsep=0.3cm
\renewcommand{\arraystretch}{0.85}
\caption{Time Delays and Uncertainties (\citealt{Bonvin2017}).}
\begin{tabular}{cccc}
\hline \hline \vspace*{-3.5mm}\\
Image Pair & $\Delta t$ & $ 1\sigma $ & $2 \sigma$ \\
\hline \vspace*{-3mm}\\
AB & -8.8 & $\pm$ 0.8 & $\pm$1.6\\
AC & -1.1 & $\pm$0.7 & $\pm$1.4 \\
AD & -13.8 & $\pm$ 0.9 & $\pm$1.8\\
\hline 
\end{tabular}
\label{time}
\end{table}	

\begin{table}
\tabcolsep=0.19cm
\renewcommand{\arraystretch}{0.85}
\caption{Flux Ratios between Images.}
\begin{tabular}{ccc}
\hline \hline \vspace*{-3.5mm}\\
Flux Ratios & Radio Emission$^*$ & $[$OIII$]$ Emission Line$^{**}$ \\
\hline \vspace*{-3mm}\\
A:B:C:D & $1.05:0.77:1.00:0.47$ & $0.97:0.98:1.00:0.66$ \\ \hline 
\end{tabular}
\label{narrow}
\tablenotetext{*}{\citealt{Jackson2015}}
\tablenotetext{**}{\ \ \citealt{Nierenberg2017}}
\end{table}	

\begin{table}
\tabcolsep=0.53cm
\renewcommand{\arraystretch}{0.85}
\caption{Lens Model Parameters (\citealt{Wong2017}).}
\begin{tabular}{ccccc}
\hline \hline \vspace*{-3.5mm}\\
Image & $\kappa$ & $\gamma$ & $\Delta \kappa^*$ & $\Delta \gamma^*$ \\
\hline \vspace*{-3mm}\\
A & 0.49 & 0.35 & 0.03 & 0.04 \\
B & 0.64 & 0.53 & 0.12 & 0.06 \\
C & 0.51 & 0.31 & 0.05 & 0.08 \\
D & 0.70 & 0.56 & 0.14 & 0.08 \\
\hline 
\end{tabular}
\label{macro}
\tablenotetext{*}{difference to our model}
\end{table}	

\section{Summary and Conclusions}
The quadruple-imaged strong gravitational lens HE 0435-1223 has four nearly identical components arranged 
symmetrically around a luminous galaxy and is an attractive target for microlensing studies because of the 
relative ease of separating intrinsic from microlensing-induced variations due to the short time delay between 
its images. Unlike most other known quadruple lens systems, photometric monitoring of this object is also 
relatively easy, because of its relatively wide image separations (\citealt{Wisotzki2002}). We used the COSMOGRAIL 
light curves of the four images of HE 0435-1223 (\citealt{Bonvin2017}, see Figure \ref{lightcurve}) to obtain the 
accretion disk size. They cover a relatively long period (13 yr), which significantly extends the time coverage of 
previous studies and provides relatively dense coverage (one observing epoch every 5 days, 884 usable points). 
Taking as reference image B, which is less affected by microlensing, and using the experimental time delays inferred by 
\citealt{Bonvin2017}, we have removed the intrinsic variability from the light curves in the overlapping region. 
Using the radio flux ratios between images determined by \citealt{Jackson2015} as a baseline for no microlensing 
magnification, we have finally obtained the microlensing light curves, A-B, C-B, and D-B. We have clearly detected 
microlensing in the images A and D of HE 0435-1223 with up to 0.6 mag (0.5 mag) in A (D). The light curve of C seems 
to be less affected by microlensing, although some changes can be seen in the first four and last two seasons of the data.\\

We have used the statistics of microlensing magnifications during the available seasons in the optical $R$-band of 
HE 0435-1223 to infer probabilistic distributions for the source size using two different methods. Using the histogram 
product of the observed and modeled microlensing histograms we have obtained a half-light radius of 
$R_{1/2}=8.4^{+11.1}_{-1.9} \sqrt{M/0.3 M_{\odot}}$ light-days. Consistent results are obtained with 
Pearson's $\chi^2$ ($R_{1/2} = 8.9^{+5.9}_{-1.2} \sqrt{M/0.3 M_{\odot}}$ light-days). In this work we 
improved the uncertainty estimations for the Pearson $\chi^2$ method to obtain self-consistent results. Our 
results are also in good agreement with previous estimates of other authors for this system
($R_{1/2} = 19^{+8}_{-6} \sqrt{M/0.3 M_{\odot}}$ light-days by \citealt{Motta2017}, 
$R_{1/2} = 5^{+4}_{-4} \sqrt{M/0.3 M_{\odot}}$ light-days by \citealt{Mosquera2011a} 
and $R_{1/2} = 6.7^{+3.0}_{-2.5} \sqrt{M/0.3 M_{\odot}}$ light-days by \citealt{Blackburne2011}).\\

Future monitoring with the The Large Synoptic Survey Telescope (LSST) will contain a large number ($\sim$8000, 
see \citealt{Oguri2010}) of light curves of gravitationally lensed quasars that will demand new techniques to compute 
the sizes of accretion disks in gravitationally lensed quasars. Here, we have explored new techniques based on 
the use of the histograms of microlensing magnifications that can be used in combination or as an alternative to 
the light curve fitting method (e.g., \citealt{Kochanek2004}).\\

\acknowledgments

We thank the anonymous referee for a thorough revision and valuable suggestions. We also thank the COSMOGRAIL collaboration for making publicly available the monitoring data of HE 0435-1223. C.F. gratefully acknowledges the financial support of a La Caixa PhD fellowship. E.M. and J.A.M are supported by the Spanish MINECO with the grants AYA2016-79014-C3-1-P and AYA2016-79014-C3-3-P. J.J.V. is supported by the Spain's Ministerio de 
Econom\'{\i}a y Competitividad and the Fondo Europeo de Desarrollo Regional (FEDER) through grant AYA2017-84897-P and AYA2014-53506-P
and by the Junta de Andaluc\'{\i}a through project FQM-108.\\

\bibliographystyle{aasjournal}
\bibliography{bib_paper}

\end{document}